\documentclass[reqno,12pt]{article}
\usepackage{amsmath}%
\usepackage{amsthm}
\usepackage{amsfonts}%
\usepackage{amssymb}%
\usepackage{graphicx}

\usepackage[comma,authoryear]{natbib}

\usepackage{parskip}
\usepackage{longtable, lscape}
\usepackage{supertabular, multicol}
\usepackage{multicol}
\usepackage{booktabs}
\usepackage{array}
\usepackage{layout}
\usepackage{setspace}
\usepackage[left=1.15in, right=1.15in, top=1.25in, bottom=1.25in]{geometry}
\usepackage{tikz,ifthen}
\usetikzlibrary{positioning}
\usetikzlibrary{calc,through,intersections,shapes.geometric}
\usepackage{ifthen}
\usepackage{geometry}
\usepackage{pgfplots} 
\allowdisplaybreaks

\newcolumntype{H}{>{\setbox0=\hbox\bgroup}c<{\egroup}@{}}
\usepackage{collcell}
\makeatletter
\newcolumntype{G}{>{\collectcell\@gobble}c<{\endcollectcell}@{}}
\makeatother
\def\eatcell#1\unskip{}
\newcolumntype{E}{>{\eatcell}c@{}}

\newtheorem{theorem}{Matrix--Tree--Theorem}[]

\newtheorem{conjecture}[theorem]{Conjecture}

\theoremstyle{definition}
\newtheorem{observation}{Remark}

\newcommand{\Keywords}[1]{\par\noindent 
{\small{\em Keywords\/}: #1}}

\newcommand{\comments}[1]{}
\newcommand{\G}{{\mathcal G}}

\newcommand{\rL}{\boldsymbol{\rho_{\Lambda}}}
\newcommand{\pT}{\boldsymbol{\psi}_{T}}

\newcommand{\diag}{\mathop{\mathrm{diag}}\nolimits}

\makeatletter
\let\mcnewpage=\newpage
\newcommand{\TrickSupertabularIntoMulticols}{%
  \renewcommand\newpage{%
    \if@firstcolumn
      \hrule width\linewidth height0pt
      \columnbreak 
    \else
      \mcnewpage 
    \fi
  }%
}
\makeatother

\renewcommand{\Lambda}{L}
\begin{document}
\title{Optimal regular graph designs}

\author{Sera Aylin Cakiroglu\footnote{Current address: CRUK London Research Institute, 44 Lincoln's Inn Fields, WC2A 3LY London, UK}\\ School of Mathematical Sciences, Queen Mary University of London,\\ Mile End Road, London E1 4NS, UK\\ Email: s.cakiroglu@qmul.ac.uk}
\date{}
\maketitle

\begin{abstract}

A typical problem in optimal design theory is finding an experimental design that is optimal with respect to some criteria in a class of designs. The most popular criteria include the $A$- and $D$-criteria. Regular graph designs  occur in many optimality results and, if the number of blocks is large enough, they are $A$- and $D$-optimal. We present the results of an exact   computer search for the best regular graph designs in  large systems for up to $20$ points, $k\leq r\leq 10$ and $r(k-1)-(v-1)\lfloor r(k-1)/(v-1)\rfloor\leq 9$. 

\Keywords{$A$-optimality; $D$-optimality; Incomplete block design; Regular graphs; Regular graph design }

\end{abstract}

\section{Introduction}
Suppose we are to design the following statistical experiment: there are $v$ treatments to be compared on a number of experimental units that can be partitioned into $b$ blocks of size $k$ with $k<v$. Typically, the blocks might differ systematically but all units in a block are assumed to be alike. For fixed $v$, $b$ and $k$, how should the treatments be allocated to the units to get as much information as possible from the available data? The latter often means the estimate of the unknown parameters with the least possible variance. If there are 
several parameters, this is a multidimensional problem and the design of the experiment can be `good' in different ways. 

We will give  the basic  definitions in the following and the reader is referred to the paper by \cite{CameronBailey}, where a good overview  and more details on the application of combinatorics can be found.
Formally, a \emph{block design} $d$ is  an assignment of $v$ treatments or varieties to a set of experimental units that have been partitioned into $b$ blocks of size $k$.  
The statistical model to analyse the data obtained from $d$ is assumed to be the linear model. For each yield $Y_{\omega j}$ of unit $\omega$  in block $j$, the model is
\[
Y_{\omega j}=\tau_{f(\omega j)}+\beta_j+\epsilon_{\omega j},
\]
where $\beta_j$ is the effect of the block $j$, $\tau_i$ is the effect of treatment $i$ and $\epsilon_{\omega j}$ are uncorrelated random errors with expectation $0$ and variance $\sigma^2$. The function $f(\omega j)$ in the above equation specifies which treatment is allocated to  unit $\omega$ in block $j$.

The design is said to be \emph{connected} if all pairwise differences $\tau_i-\tau_j$ are estimable. A design is called \emph{complete} if every treatment occurs in every block; otherwise it is called \emph{incomplete}.
If every treatment occurs at most once per block, the design is called \emph{binary} and we will always assume that all designs we consider are incomplete, binary and connected.
The \emph{replication} $r_i$ of a treatment $i$ is the total number of units that have been assigned treatment $i$. If the replications are all equal to a constant $r$, the design is called \emph{equireplicate}. 
 A well-known example for  block designs are the \emph{balanced incomplete block designs} (in short BIBDs), these are binary equireplicate incomplete block designs with replication $bk/v$, where $k<v$ and any pair of treatments is contained in exactly $\lambda$ blocks for some $\lambda>0$. BIBDs with these parameters are also called $2$-$(v,k,\lambda)$-designs.

Let $N_d$ be the $v\times b$ treatment-block incidence matrix of the design $d$, i.e. the $ij$-entry of 
$N_d$ is the number of units in block $j$ that have been assigned treatment $i$. The $ij$-entry of the product 
$N_dN_d^T$  will be denoted by $\lambda_{ij}$ and is called the \emph{concurrence} of treatments $i$ and $j$. Note that for binary designs $\lambda_{ii}=r_i$ and $\lambda_{ij}$ is the number of blocks containing both treatments $i$ and $j$. 
The \emph{information matrix} for the estimation of the treatment effects is
\[
C_d=\diag(r_1,\ldots,r_v)-\frac{1}{k}N_dN_d^T.
\]
 Since $C_d$ has row sums $0$, the all-$1$-vector is an eigenvector with eigenvalue $0$. All other eigenvalues   are   called the \emph{non-trivial  eigenvalues}. Note that if the design is connected, the rank of the information matrix is $v-1$, the eigenvalue $0$ has multiplicity $1$ and all non-trivial eigenvalues are strictly positive.  We will use another matrix associated with the design: 
The \emph{Laplacian matrix} of a binary connected  design is
\[
\Lambda_d=k \diag(r_1,\ldots,r_v)-N_dN_d^T=kC_d.
\]
Let $\rho_1(d)\geq \rho_2(d)\geq\ldots\geq\rho_{v-1}(d)\geq \rho_v(d)$ denote the eigenvalues of $\Lambda_d$.   Note that if the design is connected, then $\rho_{v-1}(d)>\rho_v(d)=0$. The eigenvalues $\rho_1(d),\ldots,\rho_{v-1}(d)$  are   called the \emph{non-trivial Laplacian eigenvalues}. 

The average variance of the set of the best linear unbiased estimators of the pairwise differences of the treatment effects is proportional to the reciprocal of the harmonic mean of the non-trivial eigenvalues of $C_d$.  The volume of the confidence ellipsoid for the estimate of $(\tau_1,\ldots,\tau_v)$ is proportional to the square root of the reciprocal of  the product of the non-trivial eigenvalues of $C_d$ (for example \cite{Shah}).  These facts give us different criteria with regards to which a design can be considered `good': A  design maximizing the harmonic mean of the non-trivial eigenvalues of $C_d$ is called \emph{A-optimal} and a design is called  \emph{D-optimal}, if it maximizes the product of the non-trivial eigenvalues of $C_d$. There are many more optimality criteria; another popular example is the $E$-criterion which is the maximization of the smallest non-trivial eigenvalue of $C_d$ and is equivalent with minimizing the largest variance of the estimators (see \cite{Shah}).

There is no general answer to the question of which design is to be chosen for given $(v,b,k)$, but there are several partial results. For example, \cite{Kiefer} proved that BIBDs are optimal with regards to a wide range of criteria, in particular the $A$- and $D$-criteria. 
But it is not clear which designs to choose if no BIBD exists. One class of designs that have been suggested to be good candidates are the regular graph designs (in short RGDs).  RGDs are equireplicate binary designs that are `close' to BIBDs  in the sense that  any pair of points occurs in either $\lambda$ or $\lambda+1$ blocks for some integer $\lambda\geq 0$. The name hints towards a connection with regular graphs which we will explain in the first section. The Laplacian matrix of an RGD $d$ with $v$ points, replication $r$ and block size $k$ can  be written as
\[
\Lambda_d=\{r(k-1)+\lambda\}{I}_v-{T}_d-\lambda{J}_v,
\]where  ${I}_v$ is the $v\times v$ identity matrix, ${J}_v$ denotes the $v\times v$ all-$1$-matrix and ${T}_d$ is a symmetric $v\times v$ $(0,1)$-matrix with $0$'s on the diagonal and exactly $r(k-1)-\lambda(v-1) $ number of $1$'s in each row and each column. Note that if $\lambda=0$, the fact that the design is connected is equivalent with the matrix ${T}_d$ being irreducible (i.e. the matrix can not be transformed into a block upper-triangular matrix by row or column permutations).

\cite{JohnMitchell}  conjectured that
if an incomplete block design is $D$-optimal (or $A$-optimal or $E$-optimal), then it is an RGD (if any RGDs exist). In the same paper, they provided a list of the best RGDs  with $v\leq 12$, $r\leq 10$ and $v\leq b$ which they found using numerical methods. 
RGDs  occur as optimal designs in many results, see for example \cite{Cheng1} and are often candidates to be used when designing an experiment in small cases. Therefore, a list of best RGDs will be very useful in practical design of experiments.

However, John and Mitchell's conjecture is not true in general (see for example \cite{RosemaryExamples}), but it holds if the number of blocks is large enough (\cite{ChengMS}). Numerical methods as used by \cite{JohnMitchell} will not extend to these large systems, but we can use exact methods capitalizing on the power of symbolic algorithms provided by Mathematica, \cite{Mathematica}.

Suppose, $\tilde{d}$ is a $2$-$(v,k,\tilde{\lambda})$-design on $v$ points and block size $k$ with  $\tilde{b}$ blocks and Laplacian matrix $\tilde{\Lambda}$.  Then for  $y\in {\mathbb N}$  the matrix $\Lambda[y]=\Lambda_d+y\Lambda_{\tilde{d}}$
is the Laplacian matrix of a design on $v$ points,  replication $r+y\tilde{\lambda}(v-1)/(k-1)$ and $b+y\tilde{b}$ blocks of size $k$. We can write $\Lambda[y]=\Lambda_d+y\Lambda_{\tilde{d}}$ with $\lambda=\lfloor r(k-1)/(v-1)\rfloor$ and  $\delta=r(k-1)-\lambda(v-1)$ as
\begin{eqnarray}\label{eqn lambda[y]}
\Lambda[y]&=&\{(r(k-1)+\lambda\}{I}_v-{T}_d-\lambda{J}_v+y\tilde{\lambda}(v{I}_v-{J}_v)\nonumber\\
&=&(\delta+vy\tilde{\lambda}+v\lambda){I}_v-(y\tilde{\lambda}+\lambda){J}_v-{T}_d\nonumber
\end{eqnarray}
and with  $x=\lambda+y\tilde{\lambda}$ we have 
\begin{eqnarray}
\Lambda[x]=(\delta+vx){I}_v-{T}_d-x{J}_v.\nonumber
\end{eqnarray}
To identify the best RGD  for large $y$, we  take a similar approach to John and Mitchell: we order all possible matrices ${T}$ according to the performance of the corresponding $\Lambda[x]$ on the $A$- and $D$-criteria for different values of $x$. In particular, we find values $x_0^A,x_0^D$ such that these orders  do not change for all $x\geq x_0^A,x_0^D$. To find the best RGDs with regards to the $A$- and $D$-criteria for given $r$ and $k$, all one has to do is to find the first matrices in the orders that produce a design. Since we are only searching among RGDs, we will say that these designs are the $A$- or $D$-best RGDs.

We would like to highlight some key differences between our approach and 	 that of \cite{JohnMitchell}.
Firstly, we have only considered $A$- and $D$-optimality since there is a list of binary connected $E$-optimal designs up to $15$ points without the restriction to RGDs due to a construction by \cite{Morgan} which can be found on www.designtheory.org. 
Secondly, to obtain a list of all the matrices ${T}$ we  use the program genreg (\cite{Mehringer}). It is  due to the restrictions of this program that we can  only consider irreducible  matrices with row sum $\delta\leq 9$. The fact that all optimal designs in \cite{JohnMitchell}  with $\delta \geq 2$ correspond to irreducible matrices makes it a reasonable pay-off to be able to handle cases with a large number of matrices. We like to note however that  RGDs corresponding to irreducible matrices might not perform better than the ones corresponding to reducible matrices, but we failed to find an example.
Also note that there is only one irreducible matrix for  $\delta=2$ or $\delta=v-1$ and we therefore exclude these cases.
And finally, as mentioned above, we are using exact methods and therefore the orderings of the matrices are valid for all $x\geq x_A,x_D$.

Our results confirm that all the designs listed in \cite{JohnMitchell} are indeed the best RGDs and moreover, that they are   the best RGDs for big $y$ (due to exact computation).  We also found $A$- and $D$-best RGDs for $v=11$, $k=r=3$ and $k=r=8$; these are two cases that are missing in \cite{JohnMitchell}. We extended the list of the best RGDs  to $v\leq 14$ and $k\leq r\leq 10$, $\delta\leq 9$ for all admissible block sizes and up to $v\leq 20$ and $k=2$.

 A list of all found RGDs not listed in \cite{JohnMitchell}  can be found in the appendix and a full list of all found designs in xml format is included in the electronic appendix. For more information on xml files see \cite{Soicher}.
The search produced examples supporting some open conjectures, which we discuss in the final section.
Moreover, we have found an example with $v=14$ and $y=0$ where the $A$-best RGD is not $D$-best and vice versa.

\section*{Acknowledgements}
The author would like to thank Peter J. Cameron and R. A. Bailey for the many conversations on optimal designs and is extremely grateful to  J.\, P. Morgan for bringing the topic to her attention and the helpful discussions about RGDs  and the arising computational difficulties. Parts of the computations were performed using the facilities at the London Research Institute, Cancer Research UK.

\section{Preliminaries}
We will denote the set of all symmetric irreducible $v\times v$ $(0,1)$-matrices with $0$'s on  the diagonal and row and column sum $\delta$  by ${\mathcal M}(v,\delta)$.

A graph $\G$ is given by a set of vertices $V(\G)$ and a set  $E(\G)$ of two-element subsets of $V(\G)$, called edges.  We allow that two vertices are joined by multiple edges but assume that there are no loops, i.e. there is no edge of the form $\{u,u\}$ for any $u\in V(\G)$. If $\G$ contains no multiple edges, we say that $\G$ is \emph{simple}. 
The number of edges $f\in E(\G)$ with $u\in f$ for a vertex $u\in V(\G)$ is called the \emph{degree} of $u$. If the degrees of all vertices of $\G$ are all equal to a constant $\delta$, then  $\G$ is called $\delta$-regular. 
A graph can be expressed as a matrix as follows: the \emph{adjacency matrix} $A(\G)$ is the symmetric $|V(\G)|\times|V(\G)|$ matrix whose $ij$-entry is the number of edges joining vertices $i$ and $j$. Note that if $\G$ is  $\delta$-regular, then any row and any column of $A(\G)$ sums to $\delta$.
If any  vertex of $\G$ can be reached from any other vertex by going along edges, then $\G$ is called \emph{connected}. In this case, $A(\G)$ is irreducible and if $\G$ is simple $A(\G)\in{\mathcal M}(v,\delta)$.

Every design $d$ with parameters $(v,b,k)$ gives  rise to a graph in a natural way   by taking the treatments  as vertices and joining any two distinct vertices $i$ and $j$ by $\lambda_{ij}$ edges, where $\lambda_{ij}$ is the concurrence of  treatments $i$ and $j$. This graph is called \emph{concurrence graph} and we will denote it by $\G_d$. Note that $\G_d$ contains no loops. The design is connected if and only if the graph $\G_d$ is connected.
If the design is binary and equireplicate with replication $r$, then   $\G_d$ is the $r(k-1)$-regular graph with adjacency matrix $A(\G)=N_dN_d^T-r(k-1){I}_v.$ The RGDs owe their name to the fact that their adjacency graph is obtained from a simple regular graph $\G$ by adding $\lambda$ edges between any two pairs of vertices. Here, $\lambda=\lfloor r(k-1)/(v-1)\rfloor$, the graph $\G$ has degree $\delta=r(k-1)-\lambda(v-1)$, its adjacency matrix  is a $(0,1)$-matrix ${T}_d$ and the Laplacian matrix of the RGD can be written as
\[\Lambda_d=\{r(k-1)+\lambda\}{I}_v-{T}_d-\lambda{J}_v.\]
That means, the Laplacian matrix of any RGD is fully determined by  the parameters $v$, $k$, $r$ and the matrix  ${T}_d$. 
In particular for $\lambda=0$, the matrix ${T}_d$ is the adjacency matrix of a connected graph and is therefore irreducible and ${T}_d\in{\mathcal M}(v,\delta)$.

Let $d$ be an RGD with Laplacian matrix $\Lambda_d$ with $v$ points and block size $k$ with  $b$ blocks and replication $r$. Further, let $\Lambda_{\tilde{d}}$ be the Laplacian matrix of a  $2$-$(v,k,\tilde{\lambda})$-design with  $\tilde{b}$ blocks, where $\tilde{\lambda}$ is minimal such that a $2$-$(v,k,\tilde{\lambda})$-design exists.  
We can write $\Lambda[y]=\Lambda_d+y\Lambda_{\tilde{d}}$ with $\lambda=\lfloor r(k-1)/(v-1)\rfloor$,  $\delta=r(k-1)-\lambda(v-1)$
and  $x=\lambda+y\tilde{\lambda}$ as 
\begin{eqnarray}\label{eqn lambda[x]}
\Lambda[x]=(\delta+vx){I}_v-{T}_d-x{J}_v.
\end{eqnarray}
Let $\psi_1\geq \psi_1\geq\ldots \geq \psi_v$ denote the eigenvalues of $\delta{I}_v-{T}_d$. Then,  $\psi_{v}=0$ since ${T}_d$ has row sum $\delta$. The non-trivial eigenvalues of $\Lambda[x]$ are $vx+\psi_1,\ldots,vx+\psi_{v-1}$. Let  $\pT$ denote the vector $(\psi_1,\ldots,\psi_{v-1})$.
The product of all non-trivial eigenvalues of $\Lambda[x]$ can be written  as a polynomial in $x$:
\begin{equation} \label{D-value y}
D(\pT,x)=\prod_{i=1}^{v-1}(vx+\psi_i)
=\sum_{j=0}^{v-1}(vx)^{v-1-j}S_j(\pT),
\end{equation}

where $S_j$ denotes the $j$th elementary symmetric polynomial, that is $S_0\equiv 1$ and for $\textbf{z}=(z_1,\ldots,z_{v-1})\in{\mathbb R}^{v-1}$
\[
S_j(\textbf{z})= \sum_{J\subseteq I,|J|=j}\prod_{i\in J} z_i,\ j=1\ldots,v-1.
\]
The harmonic mean of all non-trivial eigenvalues of $\Lambda[x]$ is
\[
A(\pT,x)=\frac{v-1}{\sum_{i=1}^{v-1}\frac{1}{vx+\psi_i}}=\frac{(v-1)D(\pT,x)}{\sum_{l=1}^{v-1} \prod_{i=1,i\not=l}^{v-1}(vx+\psi_i)}.
\]
With the relation \[\sum_{l=1}^{v-1}  S_{j;l}(\textbf{z})=(v-1-j)S_j(\textbf{z})\] (see for example \cite{Beckenbach}, p. 34) we have
\begin{eqnarray*}
\sum_{l=1}^{v-1} \prod_{i=1,i\not=l}^{v-1}(vx+\psi_i)&=&\sum_{l=1}^{v-1}\sum_{j=0}^{v-2}(vx)^{v-2-j}S_{j;l}(\pT)\\
&=&\sum_{j=0}^{v-2}(vx)^{v-2-j}(v-1-j)S_{j}(\pT).
\end{eqnarray*} Let $D_x(\pT,x)=\frac{d}{dx}D(\pT,x)$ denote the derivative of $D(\pT,x)$ in $x$, then 
\begin{eqnarray}\label{A-value as D/D'}
 A(\pT,x)&=&v(v-1)\frac{D(\pT,x)}{D_x(\pT,x)}.
\end{eqnarray}

 Note that in particular the Laplacian matrix $\Lambda_d$ of the RGD is $\Lambda_d=\Lambda[\lambda]$ and  $D(\pT,\lambda)$ and $A(\pT,\lambda)$ are the product and harmonic mean of its non-trivial eigenvalues respectively.

\section{Computation method}

For the matrix $\Lambda[x]$  to be the Laplacian matrix of an existing design, only some values for $x$ will  be admissible. But comparing designs with Laplacian matrix $\Lambda[x]$  is now reduced to comparing the values 
$D(\pT,x)$ and $A(\pT,x)$ among the matrices  ${T}\in{\mathcal M}(v,\delta)$.

The direct approach of computing the eigenvalues of $\Lambda[x]$ and computing $A(\pT,x)$ and $D(\pT,x)$ as their harmonic mean and product leads to long computation times if exact methods are used. 
It is possible however to compute $A(\pT,x)$ and $D(\pT,x)$   more efficiently: Let  $\rL=(\rho_1,\rho_2,\ldots , \rho_{v-1})$ denote the vector of the non-trivial eigenvalues of $\Lambda[x]$.
 The characteristic polynomial  $\chi(x)$ of $\Lambda[x]$  can then be written as 
\[
\chi(x)=\sum_{j=0}^v(-1)^jS_j(\rL)x^{v-j}.
\]
With equation \ref{D-value y}, we can compute  $D(\pT,x)$  with exact methods in Mathematica,  \cite{Mathematica} as the last non-zero coefficient of $\chi(x)$. Since we can express $A(\pT,x)$ as in equation \ref{A-value as D/D'}, it is enough to compute $D(\pT,x)$.

To find the best RGDs on $v$ points, we first generate all matrices in ${\mathcal M}(v,\delta)$. We do this by generating the adjacency lists of all connected $\delta$-regular graphs on $v$ vertices with the program genreg (\cite{Mehringer}) for  $\delta\leq 9$ from which we compute the  matrices $\Lambda[x]$ as given in equation \ref{eqn lambda[x]}.
As explained above, it is enough to order all matrices in ${\mathcal M}(v,\delta)$ corresponding to   the value $D(\pT,\lambda)$ and the value $A(\pT,\lambda)$  with $\lambda=\lfloor r(k-1)/(v-1)\rfloor$.

We will explain this in more detail in case of the  value $A(\pT,\lambda)$, the same procedure is applied for finding the orders with regards to $D(\pT,\lambda)$:
To find the best RGD, we want to order the matrices, say    ${\mathcal M}(v,\delta)=\{ T_1,\ldots,T_M\}$ such that
\[
A(\boldsymbol{\psi}_{T_{i}},x)>A(\boldsymbol{\psi}_{T_{i+1}},x) \text{ for all }x\geq \lambda\text{ and }i=1,\ldots,M-1
\]or, equivalently such that the polynomial 
\[P({ T_i},{T_{i+1}},x)=D(\boldsymbol{\psi}_{T_{i}},x)D_x(\boldsymbol{\psi}_{T_{i+1}},x)-D(\boldsymbol{\psi}_{T_{i+1}},x)D_x(\boldsymbol{\psi}_{T_{i}},x)\]  has no roots bigger than $ \lambda$, that is if $P(x)=0$, then $x< \lambda$.

Here, $\lambda$ still depends on the existence of RGDs corresponding to a matrix in ${\mathcal M}(v,\delta)$. 
Instead of computing the exact values for $\lambda$, we want to find the smallest integer $x^A_0$ (not depending on the existence of designs) and the order of the matrices such that 
the following two conditions are satisfied, in which case we say that the order \emph{stabilizes} for $x^A_0$.
\begin{enumerate}
 \item There exists an $x<x^A_0$  such that
\[
P({T}_{i},{T}_{i+1},x)<0\text{ for }i=1,\ldots,M-1;
\] and
\item for all $x\geq x^A_0$, \[
P({T}_{i},{T}_{i+1},x)\geq 0\text{ for }i=1,\ldots,M-1.
\]
\end{enumerate}

That means, if the order of the matrices stabilizes for $x^A_0$, we have found the order of the matrices according to $A(\pT,\lambda)$ for any admissible $\lambda\geq x^A_0$. We will say that ${T}_{i}$ has rank $i$ in this order.

We denote by  $x_0^D$  the equivalent of $x_0^A$ for the order according to $D(\pT,x)$ for ${T}\in {\mathcal M}(v,\delta)$.

We  obtain these values by first guessing a value for $x^A_0$ and $x^D_0$ and then verifying the above two properties for all matrices in ${\mathcal M}(v,\delta)$. There exist exact algorithms for the search of roots of a polynomial  in Mathematica, \cite{Mathematica}. The following table  shows the values of $x_0^A$ and $x_0^D$ for different $v$ and $\delta$.

\begin{table*}[h!]\begin{minipage}[t]{0.24\linewidth}
	\centering
		\begin{tabular}[t]{c|c|c|c}
	
  $v$ & $\delta$ & $x^A_0$& $x^D_0$  \\
	\hline
	
5&4&0&0\\
6&3&0&0\\
&4&0&0\\
7&4&0&0\\
&6&0&0\\
8&3&0&0\\
&4&0&0\\
&5&0&0\\
&6&0&0\\

	\end{tabular}\end{minipage}
\begin{minipage}[t]{0.24\linewidth}
	\centering
		\begin{tabular}[t]{c|c|c|c}
			
  $v$ & $\delta$ & $x^A_0$& $x^D_0$  \\
	\hline
	9&4&1&1\\
&6&0&0\\
	10& 3&1&1\\
& 4&1&1\\
& 5 &1&1\\
& 6&1&1\\
& 7&0&0\\
&8& 0&0\\	 
	 11 & 4&1&1 \\

\end{tabular}\end{minipage}
\begin{minipage}[t]{0.24\linewidth}
	\centering
		\begin{tabular}[t]{c|c|c|c}
			 	
  $v$ &$\delta$& $x^A_0$& $x^D_0$  \\
	
	\hline   
	 11 &6 &1&1\\
					&8&0&0\\
	 12& 3&1& 1\\
           & 4&2 &2\\ 
	 	&5&3& 3\\
		&6&3& 2\\ 
	& 7&2& 2\\
	 &8 & 1&1\\
	&9 & 0&0\\	
 
		\end{tabular}\end{minipage}
		\begin{minipage}[t]{0.24\linewidth}
	\centering
		\begin{tabular}[t]{c|c|c|c}
			 	
  $v$ &$\delta$& $x^A_0$& $x^D_0$  \\
	
	\hline    
	13 & 4& 3& 3\\
 	&6& 4&	3  \\
 &8& 5&3\\
 14 & 3& 1& 1\\
 	&4&4 &3	  \\
 &5& 6&5\\
15&4 &5 &4 \\
16&3&1&1\\
18&3&2 &1 \\
20&3&2 &2 \\
		\end{tabular}\end{minipage}
	
\end{table*}

To find the best design with replication $r$ and block size $k$, we search for the first matrix among the ordered matrices that gives rise to a block design. To do this we use the GAP package DESIGN (\cite{Soicher}).
Note that in some cases, due to either  the  large number of graphs or long computation times with GAP, we could not search among all matrices in ${\mathcal M}(v,\delta)$ and this might be a reason why we could not find any designs for some choices of $r$ and $k$.

\subsection{Table of $A$- and $D$-best RGDs in large systems}
There follows a table of the $A$- and $D$-best RGDs (for $y\geq \max\{0,(x_0^A-\lambda)/\tilde{\lambda}\}$).
We list   $\delta=r(k-1)-\lambda(v-1)$ and  the smallest $\tilde{\lambda}$ such that a $2$-$(v,k,\tilde{\lambda})$-design exists (found with GAP). 

Most designs  can be found in either \cite{Clatworthy} or \cite{CyclicDesigns}. To follow the convention as in \cite{JohnMitchell}, we write in the reference column  P.XY for a design in \cite{Clatworthy} with reference number XY and C.XY for a design in \cite{CyclicDesigns} with reference number XY. 
If the design is not in either catalogues but can be found in \cite{JohnMitchellDesignlist} we give the reference number as JM.XY and if it is cyclic we give the initial blocks.
If the design is the complement of a design in one of the  catalogue, we add an (C) to the reference number. All other designs can be found in the appendix. However, some of these designs are possible to construct with known methods, such as the ones in \cite{John67}.

\begin{multicols}{2}
\TrickSupertabularIntoMulticols

\tablehead{
 v  & k &r & $\lambda$ &$\tilde{\lambda}$ & $\delta$       & $x^A_0$&$x^D_0$&\text{Reference}\\
	 \hline  
	 }
		\label{table 2}
\begin{supertabular}[t]{cccHccHHc}
	
6 & 2 &3+5y			&0&      1  &3 &0&0   & \text{P.SR6, C.A2}\\  
	6 & 2 &4+5y			&0&      1  &4 &0&0   & \text{P.R18, C.A3}\\  
	6 & 2 &8+5y			&1&      1  &3 &0&0   & \text{P.R24}\\  
	6 & 2 &9+5y			&1&      1  &4 &0&0   & \text{P.R27}\\  

	6 & 3 &4+5y			&1&      2  &3 &0&0    &(1,2,4)(1,3,5)\\ 
	6 & 3 &7+5y			&2&      2  &4 &0&0    &\text{P.R46}\\ 
		
	6 & 3 &9+5y			&3&      2  &3 &0&0    &\text{P.R52}\\ 
	
	6 & 4 &6+10y			&3&      6  &3 &0&0  &\text{P.SR35} \\   
	6 & 4 &8+10y			&4&      6  &4 &0&0  & \text{P.R96}\\   

	7 & 2 &4+6y			&0&      1  &4  &0&0    &\text{\#1}\\ 
	7 & 2 &10+6y			&1&      1  &4  &0&0    &JM.1\\ 
	
	7 & 5 &10+15y		&6&   10     &4  &0&0    &\text{\#2}\\ 

	8 & 2 &3+7y			&0&      1  &3 &0&0    & \text{C.A7} \\ 
	8 & 2 &4+7y			&0&      1  &4  &0&0   & \text{P.SR9, C.A8} \\ 
	
	8 & 2 &5+7y			&0&      1  &5  &0&0   & JM.3\\ 
		
	8 & 2 &6+7y			&0&      1  &6  &0&0   &\text{P.R29, C.A10} \\ 
	8 & 2 &10+7y			&1&      1  &3 &0&0   &\begin{tabular}[t]{c}(1,2)(1,2)(1,3)\\(1,4)(1,5)(1,5)\end{tabular} \\ 
	
	8 & 3 &3+21y			&0&      6  &6 &0&0   & \text{P.R54, C.B3}\\  
	8 & 3 &6+21y			&1&      6  &5 &0&0   &JM.4\\  
	8 & 3 &9+21y			&2&      6  &4 &0&0   & \text{P.R58}\\  

	8 & 4 &4+7y			&1&      3  &5 &0&0   & \text{C.B6} \\  
	
	8 & 4 &6+7y			&2&      3  &4 &0&0    & \text{P.SR38}\\ 
	8 & 4 &8+7y			&3&      3  &3 &0&0   &(1,2,3,5)(1,2,3,6)\\  
	8 & 4 &9+7y			&3&      3  &6 &0&0   & \text{P.R101} \\ 

	8 & 5 &5+35y			&2&      20  &6  &0&0  & \text{P.R314, C.B3(C)} \\ 
	8 & 5 &10+35y			&5&      20  &5  &0&0  & JM.4(C)\\  
	
	8 & 6 &9+21y			&6&   15    &3  &0&0  & \text{C.A7(C)} \\  

	9 & 2 &4+8y			&0&      1  &4 &1&1   & \text{C.A11} \\  
	9 & 2 &6+8y			&0&      1  &6 &0&0    & \text{P.R34, C.A13}  \\  

	9 & 3 &3+4y			&0&      1  &6 &0&0    & \text{P.SR23} \\  

	9 & 3 &6+4y			&1&      1  &4 &0&0    & JM.8 \\ 

	9 & 3 &7+4y			&1&      1  &6 &0&0    & \text{P.R62} \\ 

	9 & 3 &10+4y			&2&      1  &4 &0&0    & JM.10\\ 
	
	9 & 4 &4+8y			&1&      3  &4  &1&1      & \text{C.B12} \\
	
	9 & 5 &5+10y			&2&      5  &4 &1&1     & \text{C.B12(C)}   \\

	9 & 6 &6+8y			&3&      5  &6  &0&0     & \text{P.SR65} \\

	10 & 2 & 3+9y                    &0 &     1&   3   &1&1  & \text{P.T2} \\

	10 & 2 & 4+9y                    &0 &     1&   4   &1&1  & \text{C.A16} \\
	10 & 2 & 5+9y                    &0 &     1&   5   &1&1  & \text{P.SR11, C.A17} \\
	10 & 2 & 6+9y                    &0 &     1&   6   &1&1  &JM.12\\
	10 & 2 & 7+9y                    &0 &     1&   7  &0&0   &JM.13\\
	10 & 2 & 8+9y                    &0 &     1&   8   &0&0  & \text{P.R36, C.A20} \\
	
	10 & 3 & 3+9y		        &0 &     2&     6  &1&1 &JM.14\\

	10 & 3 & 6+9y         &1 &     2&     3 &1&1 & \text{P.T12} \\

	10 & 4 & 4+6y        	  &1 &     2&     3  &1&1 & \text{P.T33} \\
	10 & 4 & 8+6y        	  &2 &     2&     6  &1&1 & JM.15 \\
	10 & 4 & 10+6y      	&3 &     2&     3  &1&1 & \text{P.T37} \\

	10 & 5 & 6+9y         &2 &     4&     6  &1&1 &JM.16\\
	10 & 5 & 8+9y         & 3 &     4&     5&1&1   &JM.17   \\
	10 & 5 & 10+9y       &4 &     4&     6 &1&1 &JM.19\\

	10 & 6 & 6+9y 	       &3 &     5&      3 &1&1 & \text{P.T60} \\
	
	10 & 7 & 7+21y 	       &4 &     14&      6 &1&1 &JM.14(C)\\

	11 & 2 & 4+10y                    &0 &     1&   4  &1&1& \text{C.22} \\
	11 & 2 & 6+10y                    &0 &     1&   6   &1&1 & \#3\\
	11 & 2 & 8+10y                    &0 &     1&   8 &0&0  &JM.20\\
	
	11 & 3 & 3+15y                    &0 &     3&      6 &1&1& \#4 \\

	11 & 3 & 9+15y                    &1 &     3&     8 &0&0  & JM.21\\

	11 & 4 & 8+20y                     & 2&    6&       4&1&1& \text{C.B25(C)} \\

	11 & 8 & 8+40y                     & 5&     28&       6&1&1&\# 4(C)\\
	12 & 2 & 3+11y                    &0 &     1&   3    &1&1 &JM.22\\

	12 & 2 & 4+11y                    &0 &     1&   4    &2&2& \text{C.A27} \\

	12 & 2 & 5+11y                    &0 &     1&   5   &3&3 &\# 5\\

	12 & 2 & 6+11y                    &0 &     1&   6    &3&2&\text{P.SR13,C.A29}\\

	12 & 2 & 7+11y                    &0 &     1&   7    &2&2&JM.24\\
	
	12 & 2 & 8+11y                    &0 &     1&   8    &1&1& \text{P.R38, C.A31} \\
	12 & 2 & 9+11y                    &0 &     1&   9    & 0&0& \text{P.R39, C.A32} \\
	12&  3 & 3+11y                   &0 &     2&  6    &1&1 & \# 6\\
	12 & 3 & 4+11y                   &0 &     2&   8    &1&1 & \text{P.SR26} \\
	12 & 3 & 7+11y                   &1 &     2&   3    &1&1&JM.25\\
	12 & 3 &  9+11y                   &1 &    2 &   7   &2&2& JM.27\\

	12 & 3 &  10+11y                   &1 &    2 &   9  &0 &0 & \text{P.R78, C.B34} \\

					\shrinkheight{-45em}

	12 & 4 & 3+11y                     & 0&     3&      9&0 &0&\text{P.SR41}\\

	12 & 4 & 5+11y                     & 1&     3&      4&2&2 & \text{C.B37} \\
	
	12 & 4 & 6+11y                   &   1&     3&      7&2&2&JM.28\\

	12 & 5 & 5+55y                     & 1&     20&      9& 0&0& \text{P.R145, C.B43} \\
	
	12 & 6 & 10+11y                    & 4&  5   &      6& && \text{P.SR71}\\
	
	12 & 7 & 7+77y                     & 3&    42&      9& 0& 0& \text{P.R176,}\\ &&&&&&&&\text{C.B43(C)} \\
	12 & 9 &9 +33y                   & 6&  24 &      6& 0& 0& \# 6(C) \\

	13 & 2 &  4+12y                   &0 &     1&   4   &3&3& \text{C.A35}\\
	13 & 2 &  6+12y                  &0 &     1&   6    &4&3&\#7\\
	13 & 2 &  8+12y                &0 &     1&   8    &5&3&\# 8\\
	13 & 3 &  9+6y              &1 &  1  &  6    &&&\# 9\\

	13 & 5 &  5+15y                   &1 &    5&   8   &5&3&\text{C.B58} \\
	
	\columnbreak
	14 & 2 &  3+13y                   &0 &     1&   3    &1&1& \# 10\\
	
	14 & 2 &  4+13y                  &0 &     1&   4    &4&3&\#11\\
	14 & 2 &  5+13y                &0 &     1&   5    &6&5& \# 12 \\
	
	14 & 3 &  9+39y                  &1 &     6&   5   &6&5& \# 13\\
	14 & 3 &  15+39y                  &2 &     6&   4   &4&3&\# 14\\
	
	14 & 6 &  6 +39y                 &2 &  15   &   4   &4&3&\text{C.B72}\\
	
	15 & 2 &  4+14y                   &0 &     1&   4   &5&4& \text{\#15}\\
	
	15 & 3 &  9+28y                   &1 &     1&  4    &5&4&C.B77\\
	
	16 & 2 &  3+15y                   &0 &     1&   3   &1&1 &\text{\#16}\\
	
	16 & 3 &  9+13y                       &1 &  2    &  3 &1&1 &\text{\#17} \\
	
	16 & 4 &  6+5y                       &1 &   1   &  3  &1&1& P.R118\\
	
	18 & 2 &  3+17y                   &0 &     1&  3   &2&1&\text{\#18}\\
	
	18 & 3 &  10+17y                   &1 &     2&  3  &2&1& \text{\#19}\\
	
	18 & 5 &  5 +85y                  &1 &   20  &  3  &2&1& \text{\#20}\\
	 20&2  &        3  +19y                   &0&      1   &3     &  &  &\text{\#21}\\
\end{supertabular}

\end{multicols}

\subsection{Summary and remarks}

\subsubsection*{Comparison with the results of John and Mitchell}

 For all $v\leq 12$ and $\delta\leq 5$, the best RGDs we found for $y=0$ are  isomorphic to the designs presented by John and Mitchell except for the cases  $v=11$, $k=r=3$ and $k=r=8$. All  other designs listed by John and Mitchell for $\delta \geq 2$ correspond to the same matrix in ${\mathcal M}(v,\delta)$ as the best RGDs we found. This is particularly remarkable since we made the additional restriction on all matrices to be irreducible.
In the cases $v=11$, $k=r=3$ and $k=r=8$ we found an $A$- and $D$-best RGD that John and Mitchell failed to find.  The $A$- and $D$-best RGDs correspond in both cases to the same matrix in ${\mathcal M}(11,6)$.

\subsubsection*{Order stabilization and ranks of the matrices in ${\mathcal M}(v,\delta)$}

In all cases we found that  if ${T}\in{\mathcal M}(v,\delta)$ has rank $i$ in the order corresponding to $A(\pT,x^A_0)$, then it has rank $i$ in the order corresponding to $D(\pT,x^D_0)$. The same is not true in general for the orders for $x>0$; in particular it is not true for any case where $v\geq 13$. 

The matrix with rank $1$ in both of the orders for $x=0$ remains $A$- and $D$- best with growing $x$, except for the case  $v=14$, $\delta=5$. In this case, the same is true for the order of $D(\pT,x)$, but not of $A(\pT,x)$ (see also Remark \ref{Obs 14,5}).

And finally, we have $x^A_0,x^D_0\leq\delta+1$; the worst case is $x^A_0=6$ and $x^D_0=5$ for $v=14$, $\delta=5$.

\subsubsection*{$A$- and $D$-best RGDs}

In most cases the best design was found with a matrix among the best two matrices and except for only a few cases, the matrix was among the best ten.

Except for the case $v=14$, $\delta=5$ the best RGDs for both $A(\pT,0)$ and $D(\pT,0)$ correspond to the same matrix in ${\mathcal M}(v,\delta)$. Note that since the ranks of the matrices in the orders corresponding to $A(\pT,x^A_0)$ and $D(\pT,x^D_0)$ are the same, the best RGDs for both $A(\pT,x_0^A)$ and $D(\pT,x_0^D)$ correspond to the same matrix in ${\mathcal M}(v,\delta)$ in all cases.

\bigskip

\begin{observation}{The case $v=14$, $r=5$, $k=2$:}\label{Obs 14,5}
 The best matrix in ${\mathcal M}(14,5)$ in the order for $D(\pT,0)$ (which is the same for all $x\geq 0$) is only second best for  $A(\pT,0)$, but  becomes best for  $A(\pT,x)$ where $x\geq 1$.  The best matrix for  $A(\pT,0)$ is second best for  $A(\pT,x)$ where $x\geq 1$. This means in particular that the $A$-best RGD for $v=14$, $r=5$ and $k=2$  for $y=0$ is not $D$-best and vice versa. 
\newpage
The $A$-best RGD for $y=0$ is the following design:

 \begin{tabular}[t]{ccccccccccccccccccccccccc}
1& 2& &1& 3& &1& 4& &1& 5& &1& 6& &2& 7& &
2& 8& &2& 9\\2& 10& &3& 7& &3& 8 & &3& 11& &
 4& 7& &4& 8& &4& 10 & &4& 12\\5& 7& &5& 9& &
5& 11& &5& 13& &6& 8& & 6& 10 & &6& 14& &7& 14\\
 8& 13& &9& 12& &9& 14& & 10& 11& &10& 13& &11& 12& &
 11& 14& &12& 13\\13& 14& &	3& 9& &6& 12& &\end{tabular}

 The values for this design are:
\begin{eqnarray*}
	 &A(\pT,0)&
	\simeq 68.2336883\\
	&&\\
	&D(\pT,0)&=1,627,763,046\\
	&&\\
	& A(\pT,5)&
	\simeq 1054.7827063\\
	&&\\
	& D(\pT,5)&=2,529,608,088,061,601,720,121,676.\\
	\end{eqnarray*}
	
The $A$-best RGD for $y>0$ and $D$-best RGD for $y\geq0$ is $\# 12$ in the appendix; its values are
\begin{eqnarray*}
	& A(\pT,0)
	&\simeq 68.2334019\\
	&&\\
	& D(\pT,0)&=1,627,920,000\\
	&&\\
	& A(\pT,5)
	&\simeq 1054.7827069\\
	&&\\
	& D(\pT,5)&=2,529,608,091,727,840,200,000,000.
	\end{eqnarray*}

\end{observation}

\begin{observation}
Let $K_{m,\ldots,m}$ denote the regular complete multipartite graph, that is a graph whose vertex set can be partitioned into groups of size $m$ such that any pair of vertices is joined by an edge if and only if they are in different groups. An RGD with a multipartite concurrence graph and block size $2$ is a group divisible design.

 The RGDs corresponding to the adjacency matrices of $K_{2,2,2}$, $K_{2,2,2,2}$, $K_{2,2,2,2,2}$, $K_{2,2,2,2,2,2}$, $K_{3,3,3}$ and $K_{3,3,3,3}$  stay best for all $x\geq 0$.
\end{observation} \cite{Cheng1} proved that regular complete bipartite graphs  give rise to the concurrence graphs of the unique $A$- and $D$-optimal designs for all $x\geq0$ (not necessarily only among RGDs). He extended his result to complete regular multipartite graphs among simple graphs.

\begin{observation}
We found support for the following conjecture.
\begin{conjecture}[\cite{ChengBagchi}]
 The $A$- and $D$-best RGDs with block size $2$ have an adjacency graph that  follows the pattern below.
\end{conjecture}

 Take the complete bipartite graph with parts of size $v/2-1$ and $v/2+1$ and add on the larger part the edges of a circuit on  $v/2+1$ vertices, for example for $v=10$ and $\delta=6$:
\bigskip

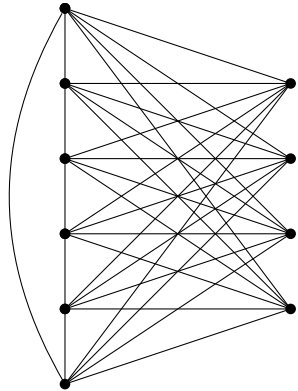
\begin{center}

\begin{tikzpicture}[every node/.style={fill,circle,inner sep=0pt,minimum size=4pt}]
\draw (0,1) node[]{}  -- (0,2) node[]{};
\draw (0,2) node[]{} -- (0,3) node[]{};
\draw (0,3) node[]{} -- (0,4) node[]{};
\draw (0,4) node[]{} -- (0,5) node[]{};
\draw (0,5) node[]{} -- (0,6) node[]{};
\draw (0,6) node[]{} edge[bend right] (0,1)node[]{};
\draw (3,2) node[]{};
\draw (3,3) node[]{};
\draw (3,4) node[]{};
\draw (3,5) node[]{};
\foreach \x in {1,2,3,4,5,6}
{
\foreach \y in {2,3,4,5}{
\draw (0,\x)--(3,\y);
}
}
\end{tikzpicture} 
\end{center}
\bigskip

These graphs are  $A$- and $D$-best for $v<14$ which was found by   \cite{JohnMitchell}. For $v=14$ we could verify the $A$- and $D$-best for this graph, too.

\end{observation}

\newpage
\appendix
\section{Optimal regular graph designs}

\underline{$\#1$ $v=7$,  $k=2$, $r=4+6y$}

\begin{tabular}{ccccccccccccccccccccccccccccccccccccccccccc}
 1& 2 & & 1& 3 & & 1& 4 & & 1& 5 & & 2& 3 & & 2& 4 & & 2& 5\\
 3& 6 & & 3& 7 & & 4& 6 & & 4& 7 & & 5& 6 & & 5& 7 & & 6& 7
\end{tabular}

\underline{$\#2$ $v=7$,  $k=5$, $r=10+15y$}

\begin{tabular}{ccccccccccccccccccccccccccccccccccccccccccc}
1& 2& 3& 4& 5 & & 1& 2& 3& 4& 6 & & 1& 2& 3& 4& 7 & & 1& 2& 3& 5& 6 \\
1& 2& 3& 5& 7 & & 1& 2& 4& 5& 6 & & 1& 2& 4& 5& 7 & & 1& 3& 4& 6& 7\\
1& 3& 5& 6& 7 & & 1& 4& 5& 6& 7 & &2& 3& 4& 6& 7 & & 2& 3& 5& 6& 7\\
 2& 4& 5& 6& 7 & & 3& 4& 5& 6& 7 
\end{tabular}

\underline{$\#3$ $v=11$,  $k=2$, $r=6+10y$}

\begin{tabular}{ccccccccccccccccccccccccccccccccccccccccccccccccccccc}
1& 2 & & 1& 3 & & 1& 4 & & 1& 5 & & 1& 6 & & 1& 7 & & 2& 3  & &
2& 6 \\ 2& 7 & & 3& 8 & & 3& 9 & & 3& 10 & & 3& 11    & &
4& 11 & & 5& 8 & & 5& 9 \\ 5& 10 & & 5& 11 & & 6& 8 & & 6& 9 & &
6& 11& & 7& 8 & & 7& 9 & & 7& 10 \\ 7& 11 & & 8& 9 & &
2& 4 & & 2& 5& & 4& 9 & & 6& 10 & & 4& 10& & 4& 8\\ 10& 11
\end{tabular}

\underline{$\#4$ $v=11$, $k=3$, $r=3+15y$}

\begin{tabular}{cccccccccccccccccccccccccccccccccccccccccccccc}
 1&2&5& &1&3&6& &1&4&7& &2&3&4& &2&8&9& &3&8&10\\
4&9&11& &5&6&11& &5&7&10& &6&9&10& &7&8&11
\end{tabular}

%
	%
	\underline{$\#5$ $v=12$, $k=2$, $r=5+11y$}
	
	\begin{tabular}{ccccccccccccccccccccccccccccccccccccccccccccccccccccccccccc}
1& 2 & & 1& 3 & & 1& 4 & & 1& 5 & & 1& 6 & & 2& 7 & & 2& 8 & & 2& 9 \\ 
3& 8 & & 3& 9 & & 3& 11 & & 4& 8& & 4& 10 & & 4& 11 & & 5& 9 & & 5& 10 \\
6& 8 & & 6& 9 & & 6& 10 & & 6& 11 & & 7& 12 & & 2& 10& & 5& 7 & &10& 12   \\
 11& 12& & 4& 7& &9& 12& &3& 7 & & 5& 11& & 8& 12
  \end{tabular}

	\underline{$\#6$ $v=12$,  $k=3$, $r=3+11y$}
	
	\begin{tabular}{ccccccccccccccccccccccccccccccccccccccccccccccccccccccccccc}
	1& 2& 3 & & 1& 4& 6 & & 1& 5& 7 & & 2& 4& 9 & & 2& 8& 10 & & 3& 5& 11\\ 3& 8& 12 & & 4& 11& 12 & & 5& 9& 10 & & 6& 7& 8 & & 6& 10& 11 & & 7& 9& 12 
	
	\end{tabular}

	\underline{$\#7$ $v=13$, $k=2$, $r=6+12y$}

 \begin{tabular}{cccccccccccccccccccccccccccccccccccccccccc}
 1& 2 & & 1& 3 & & 1& 4 & & 1& 5 & & 1& 6 & & 1& 7 & & 2& 3 & & 2& 4\\ 
2& 5 & & 2& 6 & & 2& 7 & & 3& 8 & &3& 10 & & 3& 11 & & 4& 8 & & 4& 9 \\
4& 10 & & 4& 11 & & 5& 8 & & 5& 9 & & 5& 10 & & 5& 11 & & 6& 8 & & 6& 9   \\
6& 12& &7& 10 & & 7& 11 & & 7& 12 & & 7& 13 & & 8& 12 & & 8& 13 & & 9& 12 \\
9& 13 & & 10& 12 & & 10& 13 & & 11& 12 & & 3& 9 & & 6& 13& & 11& 13
 \end{tabular}
\newpage
	{\underline{$\#8$ $v=13$, $k=2$, $r=8+12y$}
 
 \begin{tabular}{ccccccccccccccccccccccccccccccccccccccccc}
1& 2 & & 1& 3 & & 1& 4 & & 1& 5 & & 1& 6 & & 1& 7 &  & 1& 8 & &1& 9 \\ 
2& 3 & & 2& 4 & & 2& 5 & & 2& 6 & & 2& 10 & &2& 11 & & 3& 8 & & 3& 9\\
3& 10 & & 3& 11 & & 3& 12 & & 3& 13& &4& 8 & & 4& 9 & & 4& 10& &4& 11 \\
4& 13 & & 5& 8& &5& 9 & & 5& 10 & &5& 11 & & 5& 12 & & 5& 13 & & 6& 8\\
6& 9 & &6& 10 & & 6& 11 & & 6& 12& & 7& 8 & & 7& 9 & & 7& 10 & &7& 11\\
7& 12 & & 7& 13 & & 8& 10 & & 8& 12 & & 9& 11 & & 9& 13& & 2& 7& & 4& 12\\ 6& 13
 \end{tabular}

		{\underline{$\#9$ $v=13$, $k=3$, $r=9+6y$}
 
 \begin{tabular}{ccccccccccccccccccccccccccccccccccccccccc}
1&  2&  3 & & 1&  2&  3 & & 1&  4&  6 & & 1&  4&  6 &  &1&  5&  7 & & 1&  5&  7\\
1&  8&  11 & & 1&  9&  12 & &1&  10&  13 & & 2&  4&  5 & & 2&  4&  7 & & 2&  6&  8\\
 2&  8&  9 & &2&  9&  11 & & 2&  10&  12 & & 2&  10&  13 & & 3&  4&  11 & &3&  5&  6 \\
  3&  5&  9 & & 3&  7&  12 & & 3&  8&  13 & &3&  10&  11 & & 3&  12&  13 & & 4&  8&  12 \\
 4&  9&  13 & & 4&  10&  12 & & 4&  11&  13 & & 5&  8&  10 & & 5&  8&  12 & & 5&  9&  13 \\
5&  10&  11 & & 6&  7&  10 & & 6&  7&  13 & &6&  8&  11 & & 6&  9&  10 & & 6&  9&  12\\
7&  8&  13 &  7&  9&  11 & & 7&  11&  12 \end{tabular}
	
	\underline{$\#10$ $v=14$, $k=2$, $r=3+13y$} 
	
 \begin{tabular}{cccccccccccccccccccccccccccccccccccccccc}
1& 2 & & 1& 3 & & 1& 4 & & 2& 5 & & 2& 6 & & 3& 7 & & 3& 8 & & 4& 9\\
4& 10 & & 5& 11 & & 5& 12 & & 6& 13 & & 6& 14 & & 7& 11& & 7& 13 & & 8& 12 \\
8& 14 & & 9& 11 & & 9& 14 & & 10& 12 & & 10& 13
 \end{tabular}

\underline{$\#11$ $v=14$, $k=2$, $r=4+13y$}

 \begin{tabular}{cccccccccccccccccccccccccccccccccccccccc}
1& 2 & & 1& 3 & & 1& 4 & & 1& 5 & & 2& 6 & & 2& 7 & & 2& 8 \\ 3& 9 & & 3& 10 & & 4& 7 & & 4& 9 & &4& 11 & &5& 12& &
5& 13 \\ 5& 14 & & 6& 11 & &6& 12 & & 7& 10 & & 7& 13 & & 8& 9 & &
8& 12 \\ 8& 13 & & 9& 14 & &10& 12 & & 10& 14 & & 11& 13 & & 11& 14 & &
3& 6
 \end{tabular}

\underline{$\#12$ $v=14$, $k=2$, $r=5+13y$}

 \begin{tabular}[t]{ccccccccccccccccccccccccc}
1& 2 & & 1& 3 & & 1& 4 & &  1& 5 \\ 1& 6 & & 2& 7 & &
  2& 8 & &2& 9 & &2& 10 & & 3& 7 & & 3& 8  & & 3& 10 \\
 4& 7 & &4& 8 & & 4& 11 & &  4& 12& &5& 7 & &5& 9  & &
 5& 11 & & 5& 12\\6& 7 & &  6& 10  & & 6& 14 & & 8& 13 & &
  8& 14 & & 9& 13 & &9& 14 & & 10& 11 \\10& 12 & & 11& 13 & &
  11& 14 & & 12& 13& &12& 14& &3& 9& & 6& 13& &
	\end{tabular}
\newpage
\underline{$\#13$ $v=14$, $k=3$, $r=9+39y$}

 \begin{tabular}{ccccccccccccccccccccccccc}
1& 2& 11 & & 1& 2& 12 & & 1& 3& 13 & &
1& 3& 14 & & 1& 4& 9 \\ 1& 4& 10& &
1& 5& 6 & &1& 5& 8 & & 1& 6& 7& &
2& 3& 6\\ 2& 4& 13 & & 2& 5& 9 & &
2& 7& 10 & & 2& 7& 14 & & 2& 8& 9 \\ 
2& 8& 10 & & 3& 4& 8 & & 3& 5& 11& &
3& 7& 8 & & 3& 7& 9 \\ 3& 9& 10& &
3& 10& 12& &4& 5& 12 & & 4& 6& 7 & &
4& 7& 12\\ 4& 8& 11 & & 4& 11& 14 & &
5& 7& 11 & & 5& 7& 13 & & 5& 9& 14\\
5& 10& 12 & & 6& 8& 14 & & 6& 9& 13& &
6& 10& 11 & & 6& 10& 14 \\ 6& 12& 13 & &
8& 12& 13 & & 8& 13& 14 & & 9& 11& 13& &
9& 12& 14 \\10& 11& 13 & & 11& 12& 14 
	\end{tabular}

\underline{$\#14$ $v=14$, $k=3$, $r=15+39y$}

 \begin{tabular}{ccccccccccccccccccccccccccccccccccccccc}
1& 2& 9 & & 1& 2& 9 & & 1& 2& 14 & & 1& 3& 7 & & 1& 3& 7  \\ 
1& 4& 6 & & 1& 4& 6 & &1& 5& 8 & &1& 5& 8 & & 1& 5& 11 \\ 
1& 10& 11 & & 1& 10& 13 & & 1& 12& 14 & &2& 3& 12 & & 2& 3& 12   \\
2& 5& 10 & &2& 5& 10 & & 2& 6& 7 & & 2& 6& 7 & & 2& 6& 14\\
2& 7& 8  & &2& 8& 11  & &3& 4& 8 & & 3& 4& 8 & & 3& 5& 9  \\
3& 5& 9 & & 3& 6& 10 & & 3& 6& 10 & & 3& 6& 13 & & 3& 9& 11 \\
4& 5& 7 & & 4& 5& 7 & &4& 7& 9  & &4& 9& 10 & & 4& 9& 10\\ 
4& 11& 14 & &5& 6& 11 & & 5& 6& 14 & & 5& 12& 13 & & 5& 12& 13  \\
5& 13& 14 & & 6& 8& 12 & & 6& 8& 12 & & 6& 9& 12  & &6& 11& 13 \\ 
7& 9& 14 & & 7& 10& 12 & &7& 10& 12  & &7& 10& 14 & & 7& 11& 13\\
8& 9& 13  & &8& 9& 14 & & 8& 10& 13 & &8& 10& 14 & & 9& 13& 14 \\
1& 4& 12 & &2& 4& 13& & 2& 8& 11 & & 3& 10& 14& & 4& 11& 14 \\
1& 3& 13& &2& 4& 13& & 3& 11& 14  & & 4& 11& 12& & 5& 12& 14 \\ 8& 9& 12& & 10& 11& 12& & 7& 11& 13& & 6& 9& 11& & 7& 8& 13 
\end{tabular}

\underline{$\#15$ $v=15$, $k=2$, $r=4+14y$}

 \begin{tabular}{cccccccccccccccccccccccccccccccccccccccc}
1& 2 & & 1& 3 & & 1& 4 & & 1& 5 & & 2& 6 & & 2& 7 & &2& 8 \\ 
3& 6& &3& 9 &  & 3& 10 & & 4& 7& & 4& 11& & 4& 12 & & 5& 8  \\ 5& 14& &
6& 11 & & 6& 13 & & 7& 9 & & 7& 14 & & 8& 10& &8& 15 \\ 9& 13& &9& 15 & & 
 10& 12& & 10& 14 & & 11& 14 & & 11& 15 & &12& 13\\ 12& 15& &5& 13
\end{tabular}

\underline{$\#16$ $v=16$, $k=2$, $r=3+15y$}

 \begin{tabular}{cccccccccccccccccccccccccccccccccccccccc}
1& 2 & & 1& 3 & & 1& 4 & & 2& 5 & & 2& 6 & & 3& 7 & &3& 8 \\
4& 9 & & 4& 10 & & 5& 11 & & 5& 12 & & 6& 13& &6& 14 & & 7& 11 \\
7& 13 & & 8& 12 & & 8& 15 & & 9& 12& &9& 14 & & 10& 13 & & 10& 15\\
11& 16 & & 14& 16& &15& 16 
\end{tabular}

\newpage
\underline{$\#17$ $v=16$, $k=3$, $r=9+13y$}

 \begin{tabular}{cccccccccccccccccccccccccccccccccccccccc}
1&  2&  7 & & 1&  2&  8 & & 1&  3&  5 & & 1&  3&  6& &1&  4&  11\\
1&  4&  12 & & 1&  9&  10 & & 1&  13&  14& & 1&  15&  16 & & 2&  3&  15 \\
2&  4&  16 & & 2&  5&  9& &2&  5&  10 & & 2&  6&  11 & & 2&  6&  13 \\
2&  12&  14& & 3&  4&  13 & & 3&  7&  9 & & 3&  7&  16 & & 3&  8&  11 \\ 
3&  8&  14 & & 3&  10&  12 & & 4&  5&  15 & & 4&  6&  9& &4&  7&  10 \\
4&  8&  9 & & 4&  10&  14 & & 5&  6&  12 & & 5&  7&  13 & & 5&  8&  12\\
5&  11&  14 & & 5&  11&  16 & &6&  7&  14 & & 6&  8&  15 & & 6&  10&  13 \\
6&  14&  16&  &7&  8&  12 & & 7&  11&  13 & & 7&  11&  15 & & 8&  10&  15 \\  
8&  13&  16 & & 9&  11&  12 & & 9&  12&  13 & & 9&  14&  15& & 9&  14&  16 \\
10&  11&  16 & & 10&  13&  15 & & 12&  15&  16 
\end{tabular}

\underline{$\#18$ $v=18$, $k=2$, $r=3+17y$}

 \begin{tabular}{cccccccccccccccccccccccccccccccccccccccc}
 1& 2 & & 1& 3 & & 1& 4 & & 2& 5 & & 2& 6 & & 3& 7 & &3& 8 \\
4& 9 & & 4& 10 & & 5& 11 & & 5& 12 & & 6& 13& &6& 14 & & 7& 11\\
7& 13 & & 8& 15 & & 8& 16 & & 9& 12 & &9& 14 & & 10& 17 & & 10& 18 \\
11& 17 & & 12& 15 & &13& 18 & & 14& 16 & & 15& 18 & & 16& 17 \end{tabular}

\underline{$\#19$ $v=18$, $k=3$, $r=10+17y$}

 \begin{tabular}{cccccccccccccccccccccccccccccccccccccccc}
  1& 2& 7 & & 1& 2& 8 & & 1& 3& 5 & & 1& 3& 6& & 1& 4& 11 \\
	1& 4& 12 & & 1& 9& 10 & & 1& 13& 14 & &1& 15& 16 & & 1& 17& 18 \\
	2& 3& 9 & & 2& 4& 13& &2& 5& 10 & & 2& 5& 14 & & 2& 6& 11 \\
	2& 6& 12& & 2& 15& 17 & & 2& 16& 18 & & 3& 4& 14 & & 3& 7& 10\\
  3& 7& 12 & & 3& 8& 11 & & 3& 8& 13 & & 3& 15& 18& &3& 16& 17 \\
	4& 5& 6 & & 4& 7& 8 & & 4& 9& 15& &4& 9& 16 & & 4& 10& 17 \\
	4& 10& 18 & & 5& 7& 9& &5& 8& 17 & & 5& 11& 13 & & 5& 11& 18 \\
	5& 12& 15& &5& 12& 16 & & 6& 7& 17 & & 6& 8& 9 & & 6& 10& 14 \\ 
  6& 13& 16 & & 6& 13& 18 & & 6& 14& 15 & & 7& 11& 14 & & 7& 11& 16 \\
	7& 13& 15 & & 7& 13& 18 & & 8& 10& 16 & &8& 12& 15 & & 8& 14& 16\\
	8& 15& 18 & & 9& 11& 17 & &9& 12& 13 & & 9& 12& 14 & & 9& 14& 18\\
	10& 11& 15& &10& 12& 18 & & 10& 13& 17 & & 11& 12& 17 & & 14& 16& 17  \end{tabular}

	\underline{$\#20$ $v=18$, $k=5$, $r=5+85y$}
	
 \begin{tabular}{cccccccccccccccccccccccccccccccccccccccc}
  1& 2& 3& 4& 17 & & 1& 2& 5& 6& 16 & & 1& 3& 7& 8& 14 \\
	1& 12& 13& 15& 18 & & 2& 5& 8& 11& 12 & & 2& 6& 9& 13& 14  \\
	3& 5& 9& 12& 18& &3& 6& 7& 11& 15 & & 3& 8& 10& 13& 16 \\
	4& 6& 8& 13& 18 & & 4& 7& 9& 12& 16& &5& 7& 11& 13& 17 \\
	8& 9& 15& 16& 17 & & 11& 14& 16& 17& 18& & 1& 4& 9& 10& 11\\
	2& 7& 10& 15& 18& & 4& 5& 10& 14& 15& & 6& 10& 12& 14& 17\end{tabular}
	
	\newpage
	\underline{$\#21$ $v=20$, $k=2$, $r=3+19y$}
	
 \begin{tabular}{cccccccccccccccccccccccccccccccccccccccc}
  1&  2 & & 1&  3 & & 1&  4 & & 2&  5 & & 2&  6 & & 3&  7 & & 3&  8 & & 4&  9\\
  4&  10 & & 5&  11 & & 5&  12 & &6&  13 & & 6&  14 & & 7&  11 & & 7&  15 & & 8&  13 \\
   8&  16 & & 9&  12 & & 9&  17 & & 10&  18 & & 10&  19 & & 11&  18 & & 12&  16 & & 13&  20\\
   14&  15 & & 14&  19 & & 15&  17 & & 16&  19 & &17&  20 & & 18&  20
  \end{tabular}
\newpage
\addcontentsline{toc}{section}{Bibliography}

\bibliographystyle{agsm}
\bibliography{lit}


\end{document}